%% file: ms.tex
\documentclass[journal]{IEEEtran}
\IEEEoverridecommandlockouts

\usepackage[utf8]{inputenc}
\usepackage[T1]{fontenc}
\usepackage{amsmath}
\usepackage{amssymb}
\usepackage{graphicx}
\usepackage{xcolor}
\usepackage{paralist}
\usepackage{booktabs}
\usepackage[detect-all]{siunitx}
\usepackage[nolist]{acronym}
\usepackage{flushend}

\DeclareSIUnit \dBm{dBm}

\graphicspath{ {plots/}{imgs/} }

\ifCLASSOPTIONcompsoc
  \usepackage[caption=false,font=normalsize,labelfont=sf,textfont=sf]{subfig}
\else
  \usepackage[caption=false,font=footnotesize]{subfig}
\fi
\usepackage[capitalise,noabbrev]{cleveref}

\begin{document}

\begin{acronym}
	\acro{ANN}{artificial neural network}
	\acro{AoA}{angle of arrival}
	\acro{CSI}{channel state information}
	\acro{DOA}{direction of arrival}
	\acro{FFT}{fast Fourier transform}
	\acro{IATA}{International Air Transport Association}
	\acro{IFFT}{inverse fast Fourier transform}
	\acro{ISM}{industrial, scientific and medical}
	\acro{k-NN}{k-nearest-neighbor}
	\acro{LOS}{line-of-sight}
	\acro{MIMO}{multiple input multiple output}
	\acro{MLP-NN}{multilayer perceptron neural network}
	\acro{MUSIC}{multiple signal classification}
	\acro{NLOS}{non-line-of-sight}
	\acro{OFDM}{orthogonal frequency division multiplex}
	\acro{RFID}{Radio-frequency identification}
	\acro{RSS}{received signal strength}
	\acro{RMS}{root mean square}
	\acro{RSSI}{received signal strength indicator}
	\acro{RToF}{return time of flight}
	\acro{ToF}{time of flight}
	\acro{VNA}{vector network analyzer}
	\acro{WAIC}{Wireless Avionics Intra-Communication}
\end{acronym}

\title{Passive Sensing and Localization in an Aircraft Cabin Using a Wireless Communication Network}

\author{%
	\IEEEauthorblockN{%
		Fabien Geyer, %
		Thomas Multerer,
		Paulo Mendes, and
		Dominic Schupke}\\
	\IEEEauthorblockA{%
		Airbus Central Research and Technology, Munich, Germany}}
\maketitle

\input{abstract}

\input{introduction}

\input{related_work}

\input{architecture}

\input{evaluation}

\input{conclusion}

\bibliography{IEEEabrv,biblio}
\bibliographystyle{IEEEtran}

\end{document}

%% file: abstract.tex
\begin{abstract}
Advances in wireless localization techniques aiming to exploit context-dependent data has been leading to a growing interest in services able of localizing or tracking targets inside buildings with high accuracy and precision. Hence, the demand for indoor localization services has become a key prerequisite in some markets, such as in the aviation sector. 
In this context, we propose a system to passively localize and track passenger movements inside the cabin of an aircraft in a privacy preserving way using existing communication networks such as Wi-Fi or 5G.
The estimated passenger positions can be used for various automation tasks such as measurement of passenger behavior during boarding.
The paper describes a novel wireless localization system, based on Artificial Neural Networks, which passively senses the location of passengers. The position estimation is based on the observation of wireless communication signals that are already present in the environment. In this context, "passive" means that no additional devices are needed for the passengers. Experimental results show that the proposed system is able to achieve an average accuracy of \SI{12}{\centi\meter} in a challenging environment like an aircraft cabin. This accuracy seems sufficient to control passenger separation. 
\end{abstract}

%% file: introduction.tex
\section{Introduction}
\label{sec:introduction}

Localization frameworks have been investigated in the context of wireless networks. Most of the proposed localization frameworks are focused on active solutions aiming to locate targets based upon reference or active anchor nodes, which are used to obtain the relative location of the target. Such active nodes emit signals aiming to allow the localization of targets. However, such active solutions are not always suitable for several deployment scenarios and use-cases such as smart factory, as well as aeronautic and space industry.

\begin{figure}[h!]
	\includegraphics[width=\columnwidth]{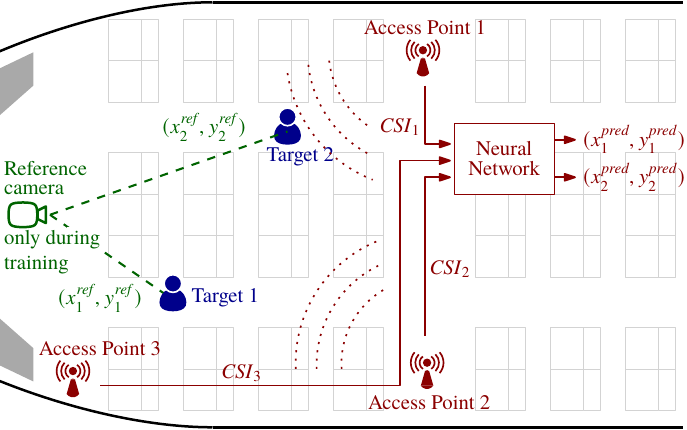}
	\caption{Overview of the on-board sensing architecture}
	\label{fig:cabin_training}
\end{figure}

This paper proposes a passive wireless localization framework, which is able to infer the location of targets (e.g. objects and people), in indoor environments as illustrated in \cref{fig:cabin_training}. It allows to monitor the location of passengers or objects in an aircraft cabin by reusing the wireless communication infrastructure of an aircraft (e.g. Wi-Fi or 5G).
Our solution is passive, meaning that no active device is required on the passenger or object to be localized.
We take advantage of the changes in the radio-waves propagation due to the presence of a target.

Having real-time information about the position of the passengers and (larger) objects in the cabin has several benefits for an aircraft operator:

\begin{itemize}
	\item Observation of passenger behavior, e.g. for queuing and boarding process analysis.
	\item Automated cabin checks for forgotten luggage.
	\item Anomaly detection with a rough position indication, e.g. missing safety equipment.
	\item Evaluation of cabin usage, e.g. free space in overhead compartments.
	\item Automated supervision and notification of physical distancing.
\end{itemize}

The above mentioned points result in a workload reduction for the cabin crew and ground personal which may lead to reduced turnaround times at the airport and hence save money. The proposed system can also be used in many other areas, like airports, train-stations or smart factories. Our system is privacy-preserving in a way so that it can localize passengers but not identify them.

We evaluate the performance of the proposed passive wireless localization system associated to the wireless infrastructure based on \SI{2.4}{\giga\hertz} Wi-Fi and \SIrange{4.2}{4.4}{\giga\hertz} \ac{WAIC} in an aircraft cabin environment.
These bands were selected since they are available worldwide, and the \ac{WAIC} band is seen as the upcoming standard band for on-board communications inside aircraft.
The radio frequency channel is known to be a challenging one because it is rich in multi-path components and obstacles. The experimental results show that the proposed wireless localization system is able to achieve an average accuracy of \SI{12}{\centi\meter} in our aircraft cabin mockup. Furthermore, we compare our numerical method to a radar signal processing approach, which achieves an average accuracy of \SI{56}{\centi\meter}. 

Additionally, we illustrate that our system is able to correctly alert wrong physical distancing behavior with an accuracy larger than \SI{90}{\percent}. This information can then used by the flight crew in order to provide guidance to passengers.%

This paper is organized as follows. We review related work on indoor localization methods in \cref{sec:relatedwork}. \cref{sec:architecture} presents our system and methods for localization. A numerical evaluation is performed in \cref{sec:evaluation}. Finally, \cref{sec:conclusion} concludes our work.

%% file: related_work.tex
\section{Related work}
\label{sec:relatedwork}

Our proposed architecture for passively localizing passengers is based on an indoor localization system.
While various solutions were already proposed for a similar case, the vast majority requires additional equipment. Besides increased costs, such solutions have further disadvantages in the aviation market: extra equipment needs to be certified and more weight is added to the aircraft. Therefore, our goal is to develop a passive localization system able to use an already existing wireless communication system in the aircraft.

In what concerns prior efforts to develop a device-free wireless localization system, \cite{Yongsen2019} and \cite{Chen2019} provide surveys of signal processing techniques and algorithms for wireless sensing. Within the research challenges, some are identified in this paper, namely multi-device cooperation and fusion of different sensors. An example is the combined usage of video and \ac{CSI} aiming to increase performance and decrease human efforts of training machine learning algorithms.

Literature shows that much of the proposed device-free indoor localization techniques are based on \ac{ToF}, \ac{RToF}, \ac{RSS}, and \ac{AoA}. In what concerns the latter, \cite{Manikanta2015} propose to compute \ac{AoA} of direct paths between the localization target and an access point by assigning values for each potential path, depending on its likelihood to be the direct path. Experimental results show that the proposed solution achieves a median accuracy of \SI{40}{\cm}. The localization error of the proposed solution depends upon the number of deployed access points. It achieves a median localization error of \SI{60}{\cm}, \SI{80}{\cm}, and \SI{190}{\cm} with five, four and three access points respectively. 

As a reference, lower errors (less than \SI{7}{\cm}) may be achieved by localization systems where targets need to use \ac{RFID} tags \cite{Kiran2015}. However, this requires the deployment of additional wireless systems operating at low frequencies (\SI{900}{\mega\hertz}).

\ac{RSS} based approaches are one of the simplest and widely used systems for indoor localization \cite{Yang2013}. Here, the signal power received at the receiver can be used to estimate the distance between transmitter and receiver.  While \ac{RSS} based approaches are simple and cost efficient, they suffer from a poor localization accuracy, especially in \ac{NLOS} conditions. This is due to additional signal attenuation resulting from obstacles, and severe fluctuations due to multipath fading.

Aiming to mitigate the limitations of \ac{RSS}-only systems, fingerprinting-based localization techniques create a better knowledge of the surrounding environment by obtaining features about the context in which the localization system is used \cite{Zafari2016}. Such features can be based on \ac{RSSI} or \ac{CSI}. Initially, different \ac{RSSI} measurements are collected during an offline phase. Once the system is deployed, the online measurements (obtained during real-time) are compared with the offline measurements to estimate the target location. 

The fingerprinting approach provides accurate positioning even in complex indoor environments without the necessity to model the wireless propagation channel. In the literature, indoor localization has been investigated for many variants of fingerprinting such as \ac{k-NN} and \ac{ANN}. Although \ac{k-NN} methods are simple to implement, this simplicity comes at the cost of relatively poor accuracy of location estimates \cite{Feng2010}.

For localization, \acp{ANN} have to be trained using \ac{RSSI} values and the corresponding target coordinates (that are obtained in the offline phase). Once the \ac{ANN} is trained, it is then used to obtain the user location based on the online \ac{RSSI} measurements. Device-free localization systems based on \acp{ANN} have been used to reach different sensing and localization goals. For instance, \cite{Shareef2007} showed that a multilayer perception \ac{ANN} combined with triangulation or trilateration techniques are able to achieve a higher accuracy and require less computational resources than Kalman filters. More recently, \acp{ANN} have been used to reach more detailed localization services, such as to estimate 3D human skeletons from RF signals \cite{Mingmin2018} based on convolutional \ac{ANN} models and camera systems for supervision. Although the proposed system detects 3D skeletons for multiple people simultaneously, it does not aim neither to localize small objects nor to tackle environments with several obstacles, such as an aircraft cabin.

In \cite{Sit2012}, a simulation for \ac{ANN}-based \ac{DOA} estimation for \ac{OFDM} \ac{MIMO} radar systems is presented. They use a \ac{MLP-NN}, with the spatial covariance matrix of the antenna array as input and the angular positions of the targets as the outputs. They found that after training of the network, the computational complexity for estimating target angles is lower than for the \ac{MUSIC} algorithm. Contrary to our approach, they use processed radar range profiles, whereas we use raw \ac{CSI} data to train the \ac{ANN}. Finally, they achieved an angular separation for two targets which were spaced less than \SI{5}{\degree} in azimuth.

To the best of our knowledge, this is the first work achieving a sufficient precision in an aircraft cabin environment using passive localization.

%% file: architecture.tex
\section{System architecture and signal processing}
\label{sec:architecture}

In this section we describe the architecture of our indoor localization system, and the signal processing which we applied. As a benchmark for the novel \ac{ANN} processing approach, we also apply a "classical" radar processing based on \ac{FFT}.

\subsection{Architecture}
Our sensing system aims to passively locate objects (e.g. passengers, luggage) and to detect environmental changes in an indoor environment, especially inside an aircraft cabin. The system is passive in the sense that the targets to be localized do not need to emit any RF signals. Hence, there is no need for any transmitter, beacon or the like on the targets. The proposed system relies only on radio measurements based on an already existing wireless communication infrastructure (e.g. based on IEEE 802.11n or 802.11ac). 

The system, illustrated in \cref{fig:cabin_training}, uses an \ac{ANN} which is trained in an offline phase before it can be used in an online phase. In the offline training phase, \ac{CSI} values are collected from the different access points. A reference system, here stereoscopic cameras, is used to annotate the \ac{CSI} values with location information ($x^{\mathit{ref}},y^{\mathit{ref}}$). Like this, the \ac{CSI} values are used as an input to the \ac{ANN}. Once the network is trained with enough \ac{CSI} values of all relevant situations, the offline phase is finished. In the online localization phase, new \ac{CSI} measurements are fed to the \ac{ANN}. Now, the \ac{ANN} estimates the position ($x^{\mathit{pred}},y^{\mathit{pred}}$) of objects in the environment.

\subsection{Signal processing}

The proposed localization system is based on multiple access points placed at different locations in the target environment. Each access point is equipped with multiple antennas which form a \ac{MIMO} array. For data transmission and localization, an \ac{OFDM} waveform (compliant to IEEE 802.11n) is used. Note that we numerically evaluate in \cref{sec:evaluation} the case where a single access point is used.

\subsubsection{Waveforms}
\label{sec:waveforms}

In this paper we use three waveforms with different bandwidths and center frequencies. The goal is to see which of the waveforms performs best in the passenger localization task. Two of them are based on Wi-Fi and one on \ac{WAIC}. In the following, $B$ is the bandwidth, $\Delta f$ is the \ac{OFDM} subcarrier spacing, $L$ is the length of the \ac{IFFT} to generate the \ac{OFDM} symbol and $N$ is the number of subcarriers.
 
First, a standard Wi-Fi signal with $B=\SI{40}{\mega\hertz}$ is generated with a $L=128$ points \ac{IFFT}. This results in a subcarrier spacing of $\Delta f = \SI{312.5}{\kilo\hertz}$. Out of the 128 possible subcarriers, only $N=114$ (-58,...,-2,2,...58) are used for data and pilots.

Second, the non-standard, Wi-Fi-like, signal with $B=\SI{100}{\mega\hertz}$ and $N=L=200$ has a subcarrier spacing of $\Delta f = \SI{500}{\kilo\hertz}$ and a frequency range of \SIrange[]{2.4}{2.5}{\giga\hertz}. 

Third, the \ac{WAIC} signal with  $B=\SI{200}{\mega\hertz}$ and $N=L=200$ has a subcarrier spacing of $\Delta f = \SI{1}{\mega\hertz}$ and its frequency range is \SIrange[]{4.2}{4.4}{\giga\hertz}. The \ac{WAIC} physical layer was chosen to be \ac{OFDM} as there is no standardization yet.

\subsubsection{Channel state information}
Unlike other approaches based on a single \ac{RSSI} value, \ac{CSI} offers the possibility to capture additional information about the physical state of the channel. \ac{CSI} describes the propagation of radio waves in a certain channel. For each frequency of interest, one complex value defines the channel. It represents the attenuation and phase shift of a signal. For \ac{OFDM} signals, \ac{CSI} values can be measured for every subcarrier. The result of one measurement is a complex-valued matrix of dimensions $N_{\mathit{TX}} \times N_{\mathit{RX}} \times N_C$, with $N_{\mathit{TX}}$ transmitting antennas, $N_{\mathit{RX}}$ receiving antennas and $N_C$ subcarriers.

The acquisition of \ac{CSI} values is available on some commodity Wi-Fi chipsets (e.g. \cite{Halperin2011}), enabling our localization system to work on off-the-shelf equipment with only some software modifications. Additionally to transmitting data packets, the access points are used to periodically measure \ac{CSI} between their antenna pairs based on the \ac{OFDM} subcarriers of the underlying communication system before being algorithmically processed for localization.

In the following, we distinguish between a "classical" method which is based on \ac{OFDM} radar and a \ac{ANN} based method.

\subsubsection{Neural network}
The \ac{ANN} makes use of features extracted from data related to radio signals aiming to fingerprint an indoor environment in the presence of different static and moving objects. This is a major advantage in relation to other passive localization systems \cite{Fadel2014} that, although having a median accuracy of \SIrange{10}{20}{\cm}, can only localize large targets (e.g. people) that are moving at any point in time.

The proposed sensing system is divided into two phases:
\begin{itemize}
	\item Gathering of \ac{CSI} values from the access points by means of a measurement agent.
	\item Localization inference by means of a predictive agent.
\end{itemize}
Both agents are installed on a wireless access point.

Indoor localization in the context of an \ac{ANN} is modeled as an approximation problem, where the task is to approximate the non-linear function that maps \ac{CSI} values (inputs) to localization coordinates (outputs) in the region of interest. In order to implement such a model, we propose a regression and classification method, whose major steps are:%
\begin{itemize}
	\item \textit{Measurement agents:}
	\begin{enumerate}
		\item Measurement of \ac{CSI} values from access points.
	\end{enumerate}
	\item \textit{Prediction agent:}
	\begin{enumerate}
		\setcounter{enumi}{1}
		\item Alignment of the measurements based on time of measurement.
		\item Extraction of the features from the \ac{CSI} values.
		\item Extraction of label and positions using an external reference localization system.
		\item Offline evaluation and training of the \ac{ANN}.
	\end{enumerate}
\end{itemize}

\begin{figure}[h!]
	\centering
	\includegraphics[width=.65\columnwidth]{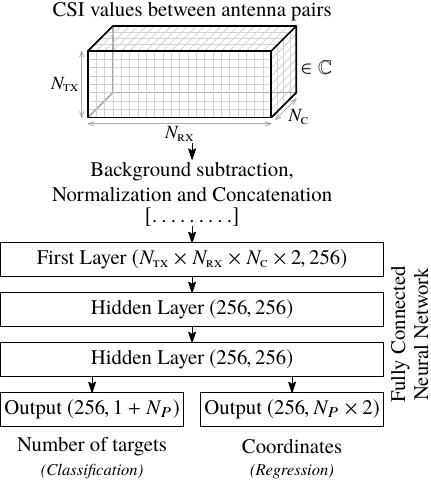}
	\caption{Overview of the \ac{ANN} architecture, with $N_P$ the maximum number of targets to localize.}
	\label{fig:nn_architecture}
\end{figure}

The evaluation process starts by gathering different samples related with raw data from the \ac{OFDM} subcarriers used by the wireless communication system. Pattern changes are detected by performing an analysis of \ac{CSI} values, allowing the proposed inference method to infer about the position of static or mobile objects. When multiple wireless access points are used, the collected data is aligned by the prediction agent according to their time of measurement in order to have a common view of the environment for a given time.
To achieve this synchronization, a dedicated network protocol for high-precision time synchronization is used (IEEE 1588).
The fully-connected architecture used by our \ac{ANN} is presented in \cref{fig:nn_architecture}.

In order to increase the target detection performance, we first apply a background subtraction step in the prediction agent, where the background of the environment (e.g. static objects) is removed. We tested various methods for this step. The simplest one is to perform multiple measurements of the empty environment at the initialization of the system, and average them in order to reduce noise. For each subsequent measurement, the averaged background is subtracted from the current measurement.
While this approach is simple to implement, it can easily be affected by drifting of the measurement equipment.
Experiments showed that the background subtraction which performed the best for the \ac{ANN} approach, was to use a Wavelet low-pass filter on the last measurements. This process removes noise from the measurements as well as movements from humans, resulting in a good background.

As a reference system for the training step, we used Intel RealSense stereoscopic depth cameras \cite{Keselman2017} which are able to provide accurate estimates of object positions. The measured \ac{RMS} error is below \SI{10}{\cm}. In order to improve the accuracy of the reference positions, visual markers, based on the ArUco library \cite{GarridoJurado2014}, are used and placed on the tracked persons and objects. This step enables us to easily collect ground-truth data by assigning ArUco tags to mock passengers and use the tracked coordinates during the training phase.
The cameras are only used during the training phase, and not deployed afterwards during inference.

The coordinates of the targets measured by the reference system are then used for training and evaluation. During the training phase, many measurements are performed where objects are moved while \ac{CSI} values are recorded.
As illustrated in \cref{fig:nn_architecture}, the \ac{ANN} provides two different outputs.
First, the number of targets, i.e. the numbers of persons currently present in the measurement environment. The \ac{ANN} is trained here following a classification approach, where the goal is to classify the scene according to the number of persons.
Second, the coordinates of the targets, by training the \ac{ANN} using a regression approach. Once the training is finalized, the evaluation phase starts, during which the position of objects is predicted using only the information coming from the wireless network architecture.
The complete architecture is illustrated and summarized in \cref{fig:cabin_training}.

\subsubsection{Classical method}
The method described here is based on \ac{OFDM} radar theory. Here, the \ac{FFT} is used to estimate range and azimuth of the targets, where range is estimated along the axis of the subcarriers, and the target angle is estimated via an \ac{FFT} across the elements of the virtual array after range processing. Two important radar parameters are the range resolution (i.e. the capability to separate two targets) of a radar system %
and the maximum unambiguous range of the radar system. %
\Cref{tab:measurement_summary} summarizes the numerical values for our system.

A uniform linear antenna array with four elements is used here. The antenna element spacing is $\lambda/2$ for the highest frequency in the band to avoid grating lobes (aliasing). As each of the antennas transmits and receives, a \ac{MIMO} configuration is given. The \ac{MIMO} principle allows to generate a virtual aperture which is not four but $V=7$ elements large. Like this, the angular resolution of the system is increased. The two-dimensional \ac{FFT} processing allows to estimate range and azimuth angles of targets. The input to the \ac{FFT} is a complex-valued matrix with the dimensions $V \times N_C$.%

Like the background subtraction in the \ac{ANN} method, sensitivity is increased by computing an image of the scattering environment from multiple measurements and then subtract it from the current measurement. This way static objects are suppressed and the actual targets become clearer. The background is formed via a median operation for every subcarrier along multiple measurements.

As detector, a simple threshold above the noise floor is used. It is slightly modified to prevent false alarms through the introduction of a certain bounding area in which the target is expected to be in the following measurement. This enables predicted coordinates consistent over time.

\subsection{Further aspects of the radar and \ac{ANN} approach}

We list here various aspects of the radar and the \ac{ANN} approach. 
First, it has to be noted, that the \ac{ANN} approach needs a training phase which incorporates as many possible scenarios as expected in the future. This is possible in many situations with recurring events which do not differ much. However, in highly variable environments, it is almost impossible to train the network and forecast all possible events. 

Second, when using pre-placed Wi-Fi access points with arbitrarily placed antennas, it is very challenging to create a model of the environment for the classical radar signal processing. The task turns into a complex multi-static radar scenario.

Third, synchronization of the measured \ac{CSI} values across all access points is necessary to create a coherent image of the scenario. Network protocols allow for a timing precision in the order of sub-microsecond. %

%% file: evaluation.tex
\section{Evaluation}
\label{sec:evaluation}

\subsection{Measurement setup}
We evaluate in this section our localization method in different environments and frequency bands. We focus in our evaluation on two relevant frequency bands for our aircraft use case, namely the ISM band, and the \ac{WAIC} band, dedicated for aircraft communications. While Wi-Fi access points, with modified firmware, can be used for measurements in the \SI{2.4}{\GHz} band, there are currently no standardized \ac{WAIC} transceivers available. Hence, we decided to use a \ac{VNA} (Rohde \& Schwarz ZNB4) for our measurements and capture the scattering parameters between four antennas placed in a linear array. With correct \ac{VNA} settings, the output is close to the one measured with a real access point. It is programmed to match the frequency band and the subcarrier spacing given above.

A summary of the measurement parameters and results are presented in \cref{tab:measurement_summary}.

\begin{table}[h!]
	\centering
	\caption{Measurement parameters and summary of results}
	\label{tab:measurement_summary}
	
	\resizebox{\columnwidth}{!}{%
	\begin{tabular}{@{} r|ccc @{}}
		\toprule
		&      \textbf{WAIC}       & \textbf{Wi-Fi \SI{100}{\MHz}} & \textbf{Wi-Fi \SI{40}{\MHz}} \\ \midrule
		Bandwidth $B$ &      \SI{200}{\MHz}      &        \SI{100}{\MHz}        &        \SI{40}{\MHz}        \\
		Sub-carrier spacing $\Delta f$ &       \SI{1}{\MHz}       &        \SI{0.5}{\MHz}        &       \SI{312.5}{\kHz}        \\
		Range resolution $\Delta r$ &    \SI{0.75}{\meter}     &       \SI{1.5}{\meter}       &      \SI{3.75}{\meter}      \\
		Maximum range $r_{max}$ &     \SI{150}{\meter}     &       \SI{300}{\meter}       &      \SI{480}{\meter}       \\ \midrule
		Indoor avg. err. ANN &      \SI{10.1}{\cm}      &        \SI{11.1}{\cm}        &       \SI{13.1}{\cm}        \\
		Radar approach &      \SI{33.3}{\cm}      &        \SI{36.8}{\cm}        &       \SI{75.2}{\cm}        \\ \midrule
		Cabin avg. err. ANN &      \SI{12.6}{\cm}      &        \SI{12.9}{\cm}        &       \SI{10.9}{\cm}        \\
		Radar approach &      \SI{63.5}{\cm}      &        \SI{65.8}{\cm}        &       \SI{64.1}{\cm}        \\ \bottomrule
	\end{tabular}}
\end{table}

The transmit power is set to the maximum of the \ac{VNA} which is \SI{12}{\dBm}.

\begin{figure}[h!]
	\includegraphics[width=\columnwidth]{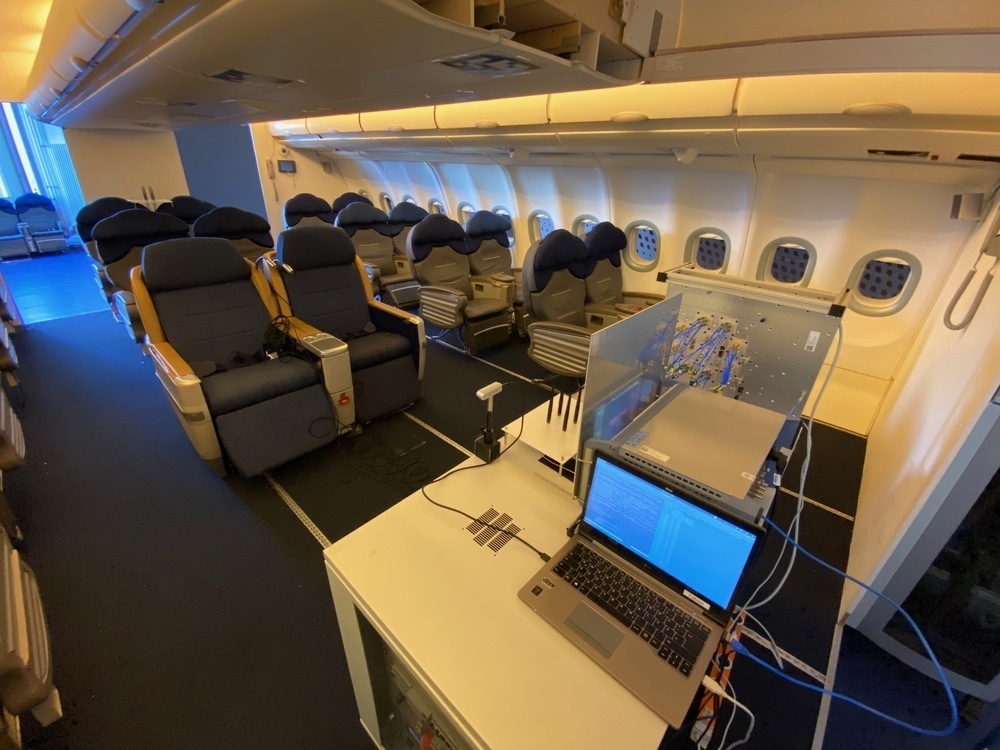}
	\caption{Cabin mockup environment used for our measurements}
	\label{fig:photo_mockup}
\end{figure}

Two different indoor environments were used for our measurements. First, we performed measurements in a empty room within a space of approximately \SI{16}{\m\squared}, where no obstacle was present. Second, a cabin mockup illustrated in \cref{fig:photo_mockup} was used. This environment is representative of an aircraft section of approximately \SI{25}{\m\squared}.
Seats may be present between the antennas and the targets to localize.

\subsection{Accuracy for locating one person}
\begin{figure}[h!]
	\centering
	\includegraphics[width=\columnwidth]{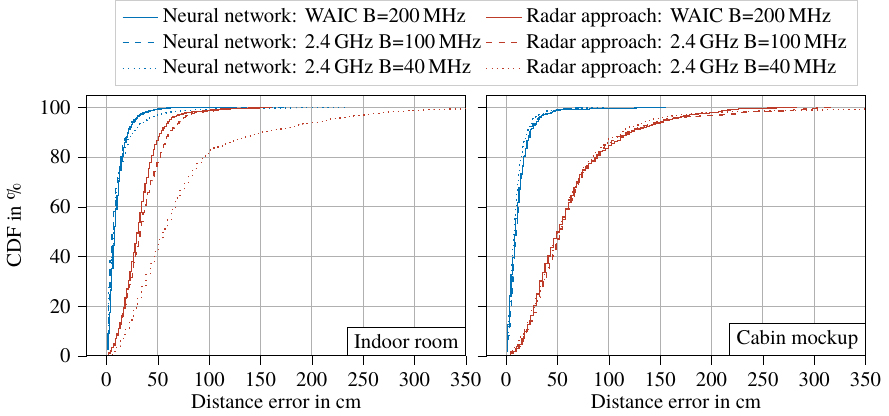}
	\caption{Accuracy of the methods in the lab and mockup environments}
	\label{fig:results_lab_mockup}
\end{figure}

In this section, measurements were performed with one person moving in the measurement environment. In order to measure the localization accuracy of our system, we use the \emph{location error} as our main metric. We define it as the Euclidean distance between the true location and the predicted location in the 2D plane. Results in our two measurements environments are presented in \cref{fig:results_lab_mockup}. In average, the \ac{ANN} was able to achieve an accuracy of \SI{11.6}{\cm}, while the radar approach achieved \SI{56.5}{\cm}.

As mentioned in \cref{sec:architecture}, the location error is inversely proportional with the bandwidth used for performing the measurement. This effect is especially relevant for the radar approach, which achieved an average location error over the two measurement environments of \SI{51.3}{\cm} (B=\SI{200}{\MHz}), \SI{48.4}{\cm} (B=\SI{100}{\MHz}), and \SI{69.7}{\cm} (B=\SI{40}{\MHz}) respectively. This effect is less relevant for the \ac{ANN} which achieved an average location error of \SI{12.7}{\cm} (B=\SI{200}{\MHz}), \SI{10.5}{\cm} (B=\SI{100}{\MHz}), and \SI{11.5}{\cm} (B=\SI{40}{\MHz}), respectively.

Finally, the measurement environment had also an important impact on the accuracy of our approaches. The \ac{ANN} achieved an average localization error of \SI{11.5}{\cm} (indoor room) and \SI{12.1}{\cm} (cabin mockup) respectively, while the radar approach achieved \SI{48.5}{\cm} (indoor room) and \SI{64.5}{\cm} (cabin mockup). This degradation in the cabin mockup is expected, mainly due to the presence of many obstacles as well as presence of more multi-paths components.

Overall, those measurements show that the \ac{ANN} is able to achieve a better accuracy compared to the radar approach. The accuracy is less influenced by the environment or the frequency band or bandwidth used.

\subsection{Accuracy for locating two persons}
In this section, measurements were performed with two persons moving in the measurement environment, with different distances between the two persons. Our goal here is to measure the distance between the two persons and decide if this distance is below a given value or not. This use case is relevant for assisting physical distancing recommendations, where the distance between two persons is measured and compared against a minimum distance requirement.

Along the localization error for each target, we define a metric for evaluating our approach called \emph{threshold accuracy}. Given a threshold distance between the two targets (e.g. a physical distancing distance), the method has an accuracy of 1 if it correctly classifies the distance between the targets as below or over this threshold, and 0 otherwise.

\begin{figure}[h!]
	\centering
	\includegraphics[width=\columnwidth]{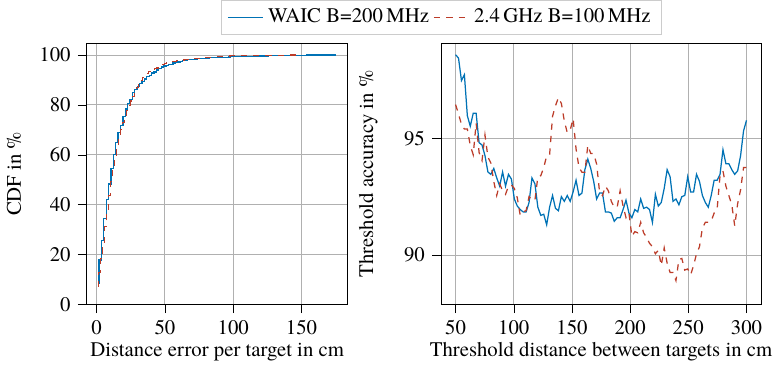}
	\caption{Accuracy of the \ac{ANN} at detecting the distance between targets}
	\label{fig:results_two_targets}
\end{figure}

Results in the cabin mockup with the \ac{ANN} are presented in \cref{fig:results_two_targets}, where it was trained at predicting the position of the two targets. The results for the distance error per target are quite similar to the ones presented in \cref{fig:results_lab_mockup}, with an average localization error per target of \SI{16.2}{\cm}.

With the threshold accuracy, we can evaluate if our approach is able to correctly emit an alert in case physical distancing is not respected over a prolonged time, i.e. if two targets are too close from each other or not. Overall, the \ac{ANN} achieved an average accuracy of approximately \SI{92}{\percent}.

%% file: conclusion.tex
\section{Conclusion}
\label{sec:conclusion}

\acresetall

In this paper we proposed and evaluated a system for localizing and tracking passenger movements inside an aircraft cabin. %
Our proposed system is reuses the existing wireless network infrastructure present aboard an aircraft with additional features such as \ac{CSI}, meaning that no additional hardware equipment is required. Our system may be used for various automation tasks dependent on the location of objects and passengers such as improving boarding speed or evaluate cabin usage.

Via experimental measurements, we compare two different approaches for analyzing the \ac{CSI} values and localizing passengers, namely a \ac{ANN} based approach and a classical radar processing approach, with different frequency ranges in the Wi-Fi and WAIC bands. As a result of the measurements performed in a cabin mockup, the \ac{ANN} is able to passively localize passengers with an average error of \SI{11.6}{\cm}, and the classical radar approach with \SI{56.5}{\cm}. We also illustrate that our system can detect passengers not respecting physical distancing with an accuracy of \SI{92}{\percent} in average.

As a next step we aim to analyze the scalability and robustness of the proposed localization scheme by exploiting the spatial feature of fingerprints from multiple adjacent access points. 
We also plan to evaluate the localization of objects and the influence of object size on accuracy.